\documentclass[aps,prl,twocolumn,superscriptaddress]{revtex4}
\usepackage{amsmath}
\usepackage{amsfonts}
\usepackage{amssymb}
\usepackage{graphicx}
\usepackage{hyperref}
\begin{document}
\title{Ferromagnetic induced Kondo effect in graphene with a magnetic impurity}
\author{Gao-Yang Li}
\affiliation{School of Physical Science and Technology,
Lanzhou University, Lanzhou 730000, China}
\author{Tie-Feng Fang}
\email[Corresponding author: ]{fangtiefeng@lzu.edu.cn}
\affiliation{School of Physical Science and Technology,
Lanzhou University, Lanzhou 730000, China}
\author{Ai-Min Guo}
\affiliation{Hunan Key Laboratory for Super-microstructure and Ultrafast Process, School of Physics and Electronics, Central South University, Changsha 410083, China}
\author{Qing-Feng Sun}
\affiliation{International Center for Quantum Materials, School of Physics, Peking University, Beijing 100871, China}
\affiliation{Collaborative Innovation Center of Quantum Matter, Beijing 100871, China}
\affiliation{CAS Center for Excellence in Topological Quantum Computation, University of Chinese Academy of Sciences, Beijing 100190, China}
\date{\today}
\begin{abstract}
We investigate the many-body effects of a magnetic adatom in ferromagnetic graphene by using the numerical renormalization group method. The nontrivial band dispersion of ferromagnetic graphene gives rise to interesting Kondo physics different from that in conventional ferromagnetic materials. For a half-filled impurity in undoped graphene, the presence of ferromagnetism can bring forth Kondo correlations, yielding two kink structures in the local spectral function near the Fermi energy. When the spin splitting of local occupations is compensated by an external magnetic field, the two Kondo kinks merge into a full Kondo resonance characterizing the fully screened ground state. Strikingly, we find the resulting Kondo temperature monotonically increases with the spin polarization of Dirac electrons, which violates the common sense that ferromagnetic bands are usually detrimental to Kondo correlations. Doped ferromagnetic graphene can behave as half metals, where its density of states at the Fermi energy linearly vanishes for one spin direction but keeps finite for the opposite direction. In this regime, we demonstrate an abnormal Kondo resonance that occurs in the first spin direction, while completely absent in the other one.
\end{abstract}
\maketitle

\section{I. introduction}
The Kondo effect \cite{Kondo1964} describes the screening of a local spin by conduction electrons, representing a paradigm of many-body correlations in condensed matter physics. Its original solution by Wilson's renormalization group method has invoked some of the most profound concepts in theoretical physics \cite{Wilson1975}. While the effect has already been studied in bulk materials for some 60 years \cite{Hewson1993}, recent two decades have seen a resurgent interest in exploring Kondo physics in artificial nanostructures \cite{Kouwenhoven2001}. Thanks to the great tunability of these nanoscale devices and their flexibility for integration with various exotic materials, many fascinating Kondo phenomena absent in bulk systems have now been revealed with unprecedented control. Of particular interest are the competition of Kondo correlations with other many-body effects (e.g., ferromagnetism \cite{Hauptmann2008} and superconductivity \cite{Franke2011, Fang2017, Fang2018}) and the screening mechanism in unconventional electronic environments \cite{Wang2019}.

The influence of itinerant electron ferromagnetism on Kondo correlations has been intensively studied in quantum dots attached to ferromagnetic electrodes \cite{Sergueev2002, Zhang2002, Lopez2003}. It was shown that in the presence of particle-hole symmetry the Kondo peak in the density of states of the dot remains unsplit even at finite lead polarizations \cite{Choi2004}. When charge fluctuations are allowed in the asymmetric case, the lead ferromagnetism causes an effective exchange field acting on the dot \cite{Martinek2003a, Martinek2005, Simon2007, Sindel2007, Matsubayashi2007}, leading to splitting and suppression of the Kondo resonance \cite{Martinek2003b, Pasupathy2004, Hamaya2007, Hofstetter2010, Li2011, Zitko2012, Wojcik2015, Weymann2017, Weymann2018}. Interestingly, this exchange field can be fully compensated by an appropriately tuned external magnetic field, thus restoring the full Kondo resonance \cite{Hauptmann2008, Martinek2003b, Hamaya2007, Simon2007, Sindel2007}. It should be emphasized that although the unsplit Kondo resonance shows up at the compensated field or in the particle-hole symmetric case, the lead ferromagnetism always plays a destructive role, which reduces the Kondo temperature \cite{Martinek2003a, Martinek2003b, Simon2007, Sindel2007, Matsubayashi2007} and hence suppresses Kondo correlations. Notably, these destructive influence of ferromagnetism were demonstrated in published studies only considering conventional ferromagnetic electrodes with smooth and featureless (if not completely flat) density of states around the Fermi energy. Detailed information on the interplay of Kondo correlations and ferromagnetism in unconventional materials with exotic band structure is still lacking.

Graphene \cite{Wallace1947, Novoselov2004, Neto2009}, with exotic Dirac-like electronic excitations, is one of such materials. It exhibits a good deal of remarkable correlated phenomena, including the fractional quantum Hall effect \cite{Du2009, Bolotin2009}, unconventional superconductivity \cite{Cao2018a}, and correlated insulating states \cite{Cao2018b}. In particular, graphene provides a perfect realization of the pseudogap Kondo problem \cite{Withoff1990, Ingersent1996, Gonzalez-Buxton1998}, when local magnetic moments are created in graphene either by adatom deposition \cite{Eelbo2013, Donati2013, Donati2014} or via point defects \cite{Yazyev2007, Wang2009, Nair2012, Ziatdinov2014, Zaminpayma2017}. The resulting Kondo physics has attracted much research interest in the last decade \cite{Cornaglia2009, Zhuang2009, Jacob2010, Vojta2010, Chen2011, Fritz2013, Mkhitaryan2013, Mastrogiuseppe2014, Lo2014, Craco2016, Ruiz-Tijerina2017, Allerdt2017, Hwang2018, Jiang2018, May2018, Zhai2019}. Graphene can even realize salient Kondo models with multiple screening channels \cite{Zhu2010, Kharitonov2013, Lee2013}, having the $SU(4)$ symmetry \cite{Wehling2010}, and exhibiting the super-Ohmic dissipation \cite{Uchoa2011}. While these studies have demonstrated the carrier doping, explicit impurity positions, and local orbital properties being crucial for characterizing the graphene Kondo system, the influence of ferromagnetism has not yet been addressed. In fact, there are several approaches \cite{Haugen2008, Wang2015, Wei2016} to induce long-range ferromagnetic order in graphene without sacrificing its excellent conductivity, e.g., placing graphene on an insulating magnetic substrate \cite{Wang2015}. Such ferromagnetic graphene provides a flexible platform for probing intriguing Kondo correlations arising from the interaction of a local moment with spin-polarized Dirac fermions.

In this paper, we shall study the Kondo physics of a magnetic adatom on ferromagnetic graphene using the numerical renormalization group (NRG) method \cite{Wilson1975, Krishna-Murthy1980, Bulla2008, Anders2005, Peters2006, Weichselbaum2007, Fang2015}. We demonstrate that unlike conventional ferromagnetic materials, the graphene ferromagnetism has constructive influence on the Kondo effect due to its exotic band dispersion. Although a particle-hole symmetric impurity in undoped graphene always keeps an unscreened local moment, Kondo correlations can be induced by the long-range ferromagnetic order present in graphene, giving rise to two kink structures in the local spectral density near the Fermi energy. By an appropriately tuned external magnetic field, the two Kondo kinks merge into a full Kondo resonance, signaling the fully screened ground state. The resulting Kondo temperature increases with the ferromagnetic exchange field and Kondo correlations are thus enhanced. Remarkably, ferromagnetic graphene can be driven into the half-metallic regime, when it is doped such that the Fermi level aligns with the Dirac point of one spin component while the hybridization of the other spin component is finite at the Fermi energy. In this regime, we demonstrate an abnormal Kondo resonance that occurs in the first spin component, but completely absent in the other one. These Kondo phenomena should be accessible by scanning tunneling spectroscopy measurements.

The remainder of the paper is organized as follows. Sec.\,II introduces the model Hamiltonian and provides some necessary details of the NRG method. Numerical results and discussion are presented in Sec.\,III, followed by a conclusion in Sec. IV.

\section{II. model and method}\label{sec:model}
The system under consideration consists of a magnetic impurity adsorbed on the ferromagnetic graphene, which is modeled by the following Hamiltonian
\begin{equation}\label{eq:Hamiltonian}
H = H_{\text{imp}} + H_{\text{g}} +H_{\text{hyb}}.
\end{equation}
Here $H_{\text{imp}}$ describes the isolated impurity,
\begin{equation}
H_{\text{imp}} = \sum_{\sigma } \varepsilon_{d}d^\dagger_\sigma d_\sigma + U d^\dagger_\uparrow d_\uparrow d^\dagger_{\downarrow}  d_{\downarrow},
\end{equation}
where $d_\sigma$ annihilates an electron with spin $\sigma=\uparrow,\,\downarrow$ and energy $\varepsilon_{d}$ in the localized magnetic orbital. $U$ is the Coulomb repulsion when the orbital is doubly occupied. In the tight-binding representation, the Hamiltonian $H_{\text{g}}$ of the ferromagnetic graphene reads \cite{Sun2010}
\begin{eqnarray}
H_{\text{g}}&=&\sum_{i,\,\sigma} (\varepsilon_0-\mu+\sigma h)(a_{i\sigma}^{\dag}a_{i\sigma}+b_{i\sigma}^{\dag}b_{i\sigma})\nonumber\\
&&-\sum_{\langle ij\rangle,\,\sigma}t(a_{i\sigma}^\dag b_{j\sigma}+\text{H.c.}),
\end{eqnarray}
where $a_{i\sigma}$ ($b_{i\sigma}$) annihilates an electron on site $i$ in sublattice $A$ ($B$) of the graphene honeycomb lattice, $\varepsilon_0$ is the Dirac-point energy of nonmagnetic graphene, $t$ is the nearest-neighbor hopping energy, and the chemical potential $\mu$ can be tuned by a gate voltage. For ferromagnetic graphene, a nonzero exchange field $h$ arises from the ferromagnetic interaction due to the proximity coupling with a magnetic insulator \cite{Wang2015}. The explicit form of the hybridization Hamiltonian $H_{\text{hyb}}$ could be very complicated, depending on the symmetry of the localized impurity orbital and its position relative to the honeycomb lattice. We consider in this work the simplest case where the impurity atom is adsorbed on the top of a carbon atom, say on site $i=0$ in sublattice $A$. In this case, the adatom can only hybridizes with this carbon atom and the specific symmetry of the localized orbital is irrelevant, yielding
\begin{equation}
H_{\text{hyb}}=\sum_{\sigma} V(a_{0\sigma}^{\dag}d_{\sigma}+\text{H.c.}),
\end{equation}
with $V$ representing the hybridization amplitude.

In the momentum space, we introduce the operators of Dirac fermions \cite{Uchoa2011},
\begin{equation}
C_{\boldsymbol{k}\sigma\alpha}=\frac{1}{\sqrt{2}}\left(a_{\boldsymbol{k}\sigma}- \alpha\frac{\phi_{\boldsymbol{k}}}{|\phi_{\boldsymbol{k}}|}b_{\boldsymbol{k}\sigma}\right),
\end{equation}
to diagonalize the graphene Hamiltonian
\begin{equation}
H_{\text{g}}=\sum_{\boldsymbol{k},\,\sigma,\,\alpha}\varepsilon_{\boldsymbol{k}\sigma\alpha} C^\dagger_{\boldsymbol{k}\sigma\alpha}C_{\boldsymbol{k}\sigma\alpha}.
\end{equation}
Here $\alpha=\pm$ labels the conduction and valence bands, $\phi_{\boldsymbol{k}}=\sum_{j=1}^{3}e^{i\boldsymbol{k}\cdot\boldsymbol{\delta}_{j}}$ with $\boldsymbol{\delta}_j$ the three vectors connecting one site with its nearest neighbors in the honeycomb lattice. Note that the presence of ferromagnetism lifts the spin degeneracy of the dispersion
\begin{equation}
\varepsilon_{\boldsymbol{k}\sigma\alpha}=\alpha t|\phi_{\boldsymbol{k}}|+\varepsilon_0-\mu+\sigma h.
\end{equation}
In this basis, the hybridization becomes
\begin{equation}
H_{\text{hyb}}=\sum_{\boldsymbol{k},\,\sigma,\,\alpha}\frac{V}{\sqrt{2N}} (C^\dagger_{\boldsymbol{k}\sigma\alpha}d_\sigma+\text{H.c.}),
\end{equation}
with $N$ the number of unit cells in the graphene.

Equations (1), (2), and (6)-(8) constitute the standard single-impurity Anderson model. Its impurity properties are fully determined by the so-called hybridization function
\begin{equation}
\Gamma_\sigma(\omega)=\frac{\pi V^2}{2N}\sum_{\boldsymbol{k},\,\alpha}\delta(\omega-\varepsilon_{\boldsymbol{k}\sigma\alpha}).
\end{equation}
Typical characteristics of $\Gamma_\sigma(\omega)$ include the presence of van Hove singularities at high energies and the low-energy linear behavior near the Dirac point. Since high-energy structures of $\Gamma_\sigma(\omega)$ are irrelevant to the Kondo physics, it is sufficient to only concentrate on its low-energy structure. Specifically, close to the Dirac point, the graphene dispersion and hence the hybridization function are linear,
\begin{eqnarray}
\varepsilon_{\boldsymbol{k}\sigma\alpha}&\simeq& \alpha\hbar v_{F}|\boldsymbol{k}| + \varepsilon_0-\mu+\sigma h,\\
\Gamma_\sigma(\omega)&\simeq& \frac{V^2\Omega_0}{N\hbar^2 v_F^2}|\omega-\varepsilon_0+\mu-\sigma h|,
\end{eqnarray}
measuring $\boldsymbol{k}$ with respect to the Dirac-point momentum. Here $v_F$ is the Fermi velocity and $\Omega_0$ is the area of graphene unit cell. For notation simplicity, a hybridization-function prefactor, $\Gamma_0\equiv\frac{V^2\Omega_0D}{N\hbar^2v^2_F}$ with $D$ the half bandwidth of Dirac fermions, is used hereafter. It is apparent that in ferromagnetic graphene, the massless Dirac fermions, characterized by the linear dispersion of Eq.\,(10), are spin polarized due to the long-range ferromagnetic ordering. Consequently, the hybridization function Eq.\,(11), which fully accounts for the influence of the Dirac-fermion bath on the impurity, is also spin polarized. It is our aim in this paper to explore the graphene Kondo physics subject to this spin polarization of Dirac fermions.

\begin{figure}
\begin{center}
\includegraphics[width=1.0\columnwidth]{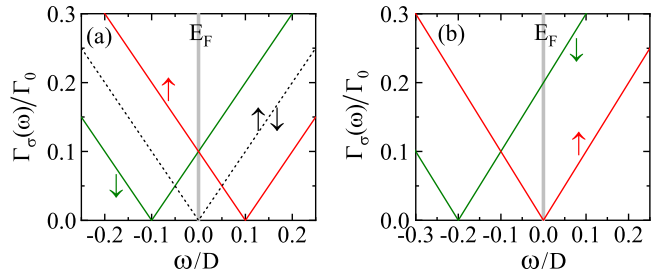}
\caption{Spin-resolved hybridization function $\Gamma_\sigma(\omega)$ at $\mu=\varepsilon_0$ (a) and $\mu=\varepsilon_0+h$ (b). Solid and dashed thin lines represent the hybridization of a top-site impurity with ferromagnetic ($h/D=0.1$) and nonmagnetic ($h/D=0$) graphene, respectively. The thick gray lines indicate the Fermi energy $E_F$.}
\end{center}
\end{figure}

The nontrivial spin splitting of the hybridization function, as shown in Fig.\,1, suggests that the Kondo physics will exhibit interesting modulations under variation of the ferromagnetic exchange field $h$ and/or the chemical potential $\mu$. It is well established \cite{Fritz2013} that in the presence of particle-hole symmetry, no Kondo screening is possible for impurities in neutral ($\mu=\varepsilon_0$) and nonmagnetic ($h=0$) graphene for which $\Gamma_\sigma(\omega)$ vanishes as $|\omega|$ at the Fermi energy $E_F=0$ [the dashed line in Fig.\,1(a)]. As soon as the ferromagnetism is turned on, the hybridization function at the Fermi energy acquires a finite value $\Gamma_0|h|/D$ for both spin species [the intersection of the two solid lines in Fig.\,1(a)], in favour of developing the Kondo screening as in normal metals. On the other hand, the spin-splitting of the hybridization function away from the Fermi energy is obviously unfavourable to the Kondo effect. We expect intriguing Kondo phenomena arising due to these two conflicting consequences of the ferromagnetism in graphene. Remarkably, when the chemical potential $\mu$ is tuned to the spin-dependent Dirac-point energy $\varepsilon_0+\sigma h$, say $\mu=\varepsilon_0+h$, the ferromagnetic graphene behaves like a Dirac half metal: the Fermi level aligns with the spin-$\uparrow$ Dirac point while the spin-$\downarrow$ hybridization function is finite at the Fermi energy [Fig.\,1(b)]. Can Dirac fermions in this exotic half-metal regime still screen the local spin? If can, what is the manifestation of the resultant Kondo effect?

We shall address these issues by using the NRG method \cite{Wilson1975, Krishna-Murthy1980, Bulla2008} based on the full density matrix algorithm \cite{Anders2005, Peters2006, Weichselbaum2007, Fang2015}. The full density-matrix NRG \cite{Weichselbaum2007} is one of the most powerful methods for quantum impurity systems. It iteratively diagonalizes the Hamiltonian and yields an approximate but complete set of eigenstates. This complete basis set \cite{Anders2005} can be used to calculate dynamical properties \cite{Peters2006} such as the impurity spectral function, as well as thermodynamic properties \cite{Fang2015} such as the impurity entropy. What follows are the numerical results obtained in the units of $D=k_B=\hbar=1$. NRG calculations are performed by using a discretization parameter $\Lambda=1.8$ for dynamical properties and $\Lambda=1.8\sim2.5$ for thermodynamic quantities, and retaining $M_K=512\sim 1024$ states per iteration. Discrete spectral data is smoothened based on the log-Gaussian kernel proposed in Ref.\,\cite{Weichselbaum2007} with a broadening parameter $\alpha=0.4\sim0.5$. Results are $z$ averaged over $N_z=1\sim4$ calculations. Throughout this work, we fix the Dirac-point energy $\varepsilon_0=0$ and the local Coulomb interaction $U=0.2$. The temperature $T$ is set to zero for all results, unless indicated otherwise.

\section{III. Numerical results and discussion}\label{sec:results}
\subsection{A. Charge-neutral nonmagnetic graphene}
We first revisit the Kondo physics of a magnetic adatom on the nonmagnetic ($h=0$) graphene at charge neutrality ($\mu=0$). Figure 2(a) displays the resulting phase diagram in the plane spanned by the impurity particle-hole asymmetry $\delta=\varepsilon_d+U/2$ and the hybridization $\Gamma_0$. The phase boundary is determined according to the zero-temperature values of the impurity contribution to entropy \cite{Bulla2008}, $S_{\text{imp}}(T)=S(T)-S_0(T)$. Here
\begin{equation}
S(T)=\beta\langle H\rangle+\ln[\textrm{Tr}(e^{-\beta H})]
\end{equation}
and
\begin{equation}
S_0(T)=\beta\langle H_\text{g}\rangle+\ln[\text{Tr}(e^{-\beta H_\text{g}})],
\end{equation}
with $\beta=1/T$, are the entropy of the system with and without the impurity, respectively. The central (green) region of Fig.\,2(a) represents the local moment phase, characterized by a residual entropy $S_\text{imp}(0)=\ln 2$ [Fig.\,2(b)]. In this doublet phase, the impurity moment is asymptotically decoupled from the graphene and behaves like a free local moment. On the other hand, in the two side (orange) regions of Fig.\,2(a), the impurity entropy goes to zero, $S_\text{imp}(0)=0$ [Fig.\,2(b)], indicating a fully screened singlet ground state that is described by the asymmetric strong-coupling fixed point. The separatrix between the singlet and doublet phases is actually a many-body level crossing, which can also be determined by the discontinuities of the ground-state energy as a function of $\delta$ and $\Gamma_0$. In the screened singlet phase, Kondo correlations can develop when charge fluctuations are suppressed. But there is no sharp boundary between the Kondo and mixed-valence regimes. It is instead a smooth crossover as the impurity level, $\varepsilon_d$ or $\varepsilon_d+U$, approaches the Fermi level.

\begin{figure}[t]
\begin{center}
\includegraphics[width=1.0\columnwidth]{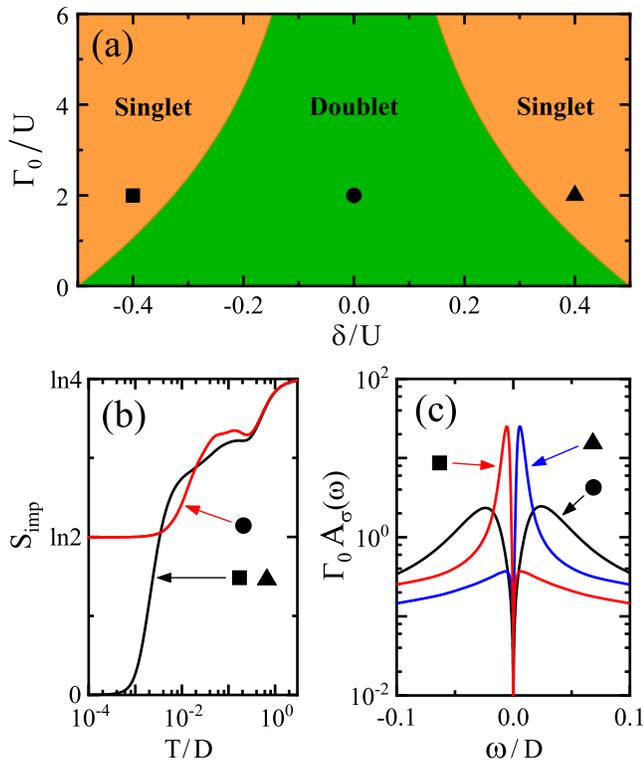}
\caption{(a) Impurity phase diagram for charge-neutral nonmagnetic graphene in the plane spanned by the impurity asymmetry $\delta$ and the hybridization $\Gamma_0$. (b) Impurity entropy $S_\text{imp}(T)$ and (c) spectral function $A_\sigma(\omega)$, for the parameter set $\{\delta/U,\,\Gamma_0/U\}=\{-0.4,\,2\}\left({\scriptscriptstyle\blacksquare}\right)$, $\{0,\,2\}(${\large$\bullet$}$)$, and $\{0.4,\,2\}(\blacktriangle)$ marked in (a).}
\end{center}
\end{figure}

A striking feature of the Kondo effect in neutral nonmagnetic graphene is that the impurity spectral function \cite{Bulla2008}, $A_\sigma(\omega)=-\frac{1}{\pi}\text{Im}G^\text{ret}_\sigma(\omega)$ with
\begin{equation}
G^\text{ret}_\sigma(\omega)=-i\int^\infty_{-\infty}\text{d}t\,\,e^{i\omega t}\,\theta(t)\left\langle\left\{d_\sigma(t),\,d_\sigma^\dagger\right\}\right\rangle,
\end{equation}
is not peaked at the Fermi energy, exhibiting no characteristic Kondo resonance. In fact, as shown in Fig.\,2(c), both in the Kondo and local-moment regimes, the impurity spectral density vanishes linearly at the Fermi energy, but featuring broad peaks away from $E_F$. It seems that there is no clear-cut spectral signature to distinguish these two phases. All the features discussed in this subsection are more or less understood in the literature (see Ref.\,\cite{Fritz2013}, for a review).

\subsection{B. Ferromagnetic graphene at charge neutrality}
Now we turn to investigate the effect of ferromagnetic graphene on the impurity atom. In Sec.\,III B, we fix the chemical potential $\mu=0$ to obtain results for undoped graphene, while Sec.\,III C presents results in the Dirac half-metal regime reached by tuning $\mu$. In both subsections, we always consider the impurity being particle-hole symmetric ($\delta=0$).

Starting from the local moment regime at zero exchange field ($h=0$), Fig.\,3(a) shows the impurity spectral feature $A(\varepsilon)=A_\uparrow(\omega)+A_\downarrow(\omega)$ with increasing $h$. Besides the broadening of the Hubbard bands and the increasing of spectral density at the Fermi energy [inset of Fig.\,3(b)], the most striking phenomenon induced by the exchange field is the appearance of two kink (or shoulder) structures near the Fermi energy. Note that as $h$ increases, the two kinks move away from each other but seem more evident. We attribute them to the Kondo effect arising from many-body correlations between the impurity and spin-polarized Dirac electrons in graphene, as demonstrated in the following.

\begin{figure}[t]
\begin{center}
\includegraphics[width=0.95\columnwidth]{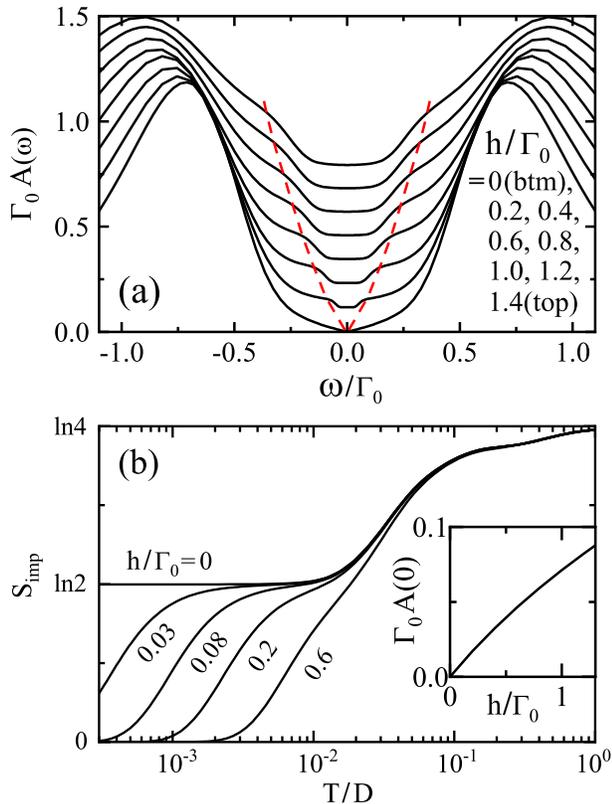}
\caption{(a) Impurity spectral density $A(\omega)$ and (b) impurity entropy $S_{\text{imp}}(T)$ for different ferromagnetic exchange field $h$. Parameters used are $\mu=0$, $\Gamma_0=0.1$, and $\delta=0$. The curves in (a) are offset for clarity, and the dashed lines guide the evolution of kink structure in $A(\omega)$. The inset of (b) shows the spectral density at the Fermi energy as a function of $h$.}
\end{center}
\end{figure}

The corresponding evolution of the impurity entropy $S_{\textrm{imp}}(T)$ is shown in Fig.\,3(b), which illustrates a suppression of $S_{\textrm{imp}}(T)$ by finite $h$ at low temperatures. Unlike in nonmagnetic metals, this suppression of the impurity entropy is not an unambiguous signature of the fully screened Kondo state. The latter in fact cannot develop in the presence of spin asymmetry caused by the exchange field. In other words, the impurity entropy is not a good quantity for characterizing Kondo physics in ferromagnetic graphene.

In order to elaborate the Kondo nature of the kink structure found in Fig.\,3(a), we investigate in detail the kink position as a function of the exchange filed $h$. Since it is difficult to precisely determine the kink position, the following strategy is adopted.  We assume that the kink structure is formed by the superposition
\begin{equation}
F(\omega)=F_1(\omega)+F_2(\omega)
\end{equation}
of the Kondo resonance peak \cite{Frota1992}
\begin{equation}
F_1(\omega)=c_1 \text{Re}\bigg[\bigg(\frac{i c_3}{\omega-c_2+i c_3}\bigg)^{\frac{1}{2}}\bigg]
\end{equation}
and a linear spectrum $F_2(\omega)=c_4+c_5|\omega|$. Here parameters $c_1\sim c_5$ are determined by using Eq.\,(15) to fit the spin-resolved NRG spectral data [Figs.\,4(a) and 4(b)]. In particular, we take the fitting parameter $c_2$, which gives the position of the Kondo resonance peak, as the kink position. The resulting dependence of the kink position on the field $h$ is given in Fig.\,4(c). On the other hand, perturbative scaling analysis \cite{Martinek2005, Sindel2007} indicates that the ferromagnetism of Dirac electrons can induce a local exchange field acting on the impurity, which in turn leads to a spin-dependent renormalization $\widetilde{\varepsilon}_{d\sigma}$ and a spin splitting $\Delta=|\widetilde{\varepsilon}_{d\uparrow}-\widetilde{\varepsilon}_{d\downarrow}|$ of the impurity level $\varepsilon_d$:
\begin{equation}
\widetilde{\varepsilon}_{d\sigma}=\varepsilon_d-\frac{1}{\pi}\int \textrm{d}\omega\,\bigg\{ \frac{\Gamma_{\sigma}(\omega) [1-f(\omega)]}{\omega-\varepsilon_{d}} +\frac{\Gamma_{\bar{\sigma}}(\omega) f(\omega)}{\varepsilon_{d}+ U -\omega} \bigg\},
\end{equation}
where $f(\omega)$ is the Fermi distribution function. As shown in Fig.\,4(c), the local spin splitting $\pm\Delta$ calculated from Eq.\,(17) are in good agreement with the kink position calculated by the NRG method. This implies that the two kink structures appearing in the spectral function are actually two split Kondo peaks distorted by graphene's linear dispersion.

\begin{figure}[t]
\includegraphics[width=\columnwidth]{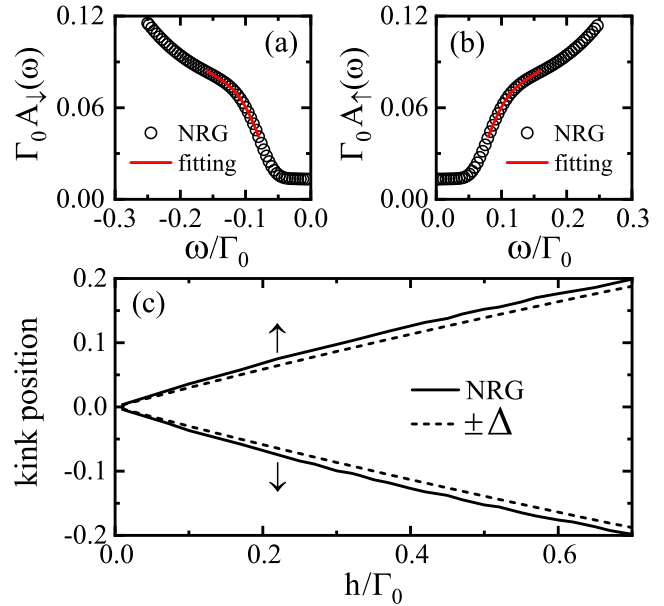}
\caption{(a) and (b) NRG results for the Kondo shoulder in the spin-resolved spectral function $A_\sigma(\omega)$ at $h/\Gamma_0=0.32$, fitted by Eq.\,(15) in the main text. (c) NRG results for the position of spin-resolved Kondo shoulder as a function of the ferromagnetic exchange field $h$, in comparison with the effective local-level splitting $\pm\Delta $ [from Eq.\,(17) in the main text]. Parameters used are $\mu=0$, $\Gamma_0=0.1$, and $\delta=0$.}
\end{figure}

Another feature that reveals the Kondo nature of the kink structure is its magnetic field dependence. We apply a local magnetic field $B$ on the impurity atom. This adds the Zeeman energy, $B(n_\uparrow-n_\downarrow)$ with $n_\sigma=d^\dagger_\sigma d_\sigma$, to the Hamiltonian (1). Figures 5(a) and 5(b) present the evolution of Kondo kink with the magnetic field. While the magnetic field $B$ parallel to the exchange field $h$ simply suppresses the Kondo kinks [Fig.\,5(b)], the situation is more interesting when $B$ is antiparallel to $h$ [Fig.\,5(a)]. As the antiparallel magnetic field increases, the two kinks first approach to each other and deform into two peaks. The two peaks then merge into a single Kondo resonance at a particular field $B=B_c$ [see the red curve in Fig.\,5(a)]. Increasing further the magnetic field, the resonance peak splits again and eventually fades away. Since at $B=B_c$ the spin asymmetry in the local occupancies also vanishes [see the point $n_\uparrow=n_\downarrow=0.5$ in Fig.\,5(c)], we actually obtain a fully-screened ground state and the strong-coupling Kondo limit is reached at $B_c$. This compensation effect of the external magnetic field has already been observed for Kondo effect in conventional ferromagnetic materials \cite{Hauptmann2008, Martinek2003b, Hamaya2007, Simon2007, Sindel2007}. Unlike the previous cases, the compensation field in our case needs to be extremely finely tuned because the step in the occupancy $n_\sigma(B)$ as a function of $B$ is very steep [Fig.\,5(c)] due to the exotic linear spectrum of ferromagnetic graphene.

\begin{figure}[t]
\includegraphics[width=\columnwidth]{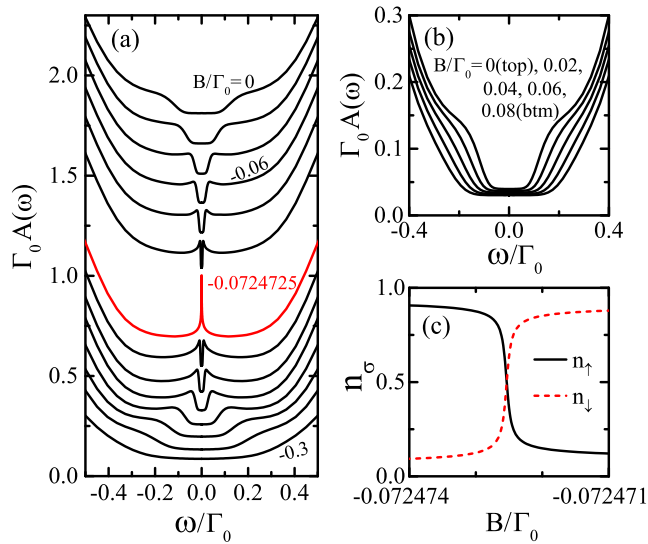}
\caption{(a),(b) Evolution of the Kondo shoulder in the spectral function $A(\omega)$ with the variation of external magnetic field $B$. (c) Spin-resolved local occupation $n_\sigma$ as a function of $B$. Note that the magnetic field at which $n_\uparrow=n_\downarrow=0.5$ in (c) is exactly the same magnetic field where the two Kondo shoulders merge into a single Kondo peak [the red curve in (a)]. Parameters used are $h/\Gamma_0=0.5$, $\mu=0$, $\Gamma_0=0.1$, and $\delta=0$. }
\end{figure}

In the presence of external magnetic field $B$, the renormalization Eq.\,(17) of the impurity level within the perturbative scaling theory \cite{Martinek2005, Sindel2007} should be modified in the following way: in the right hand side of Eq.\,(17), the first two $\varepsilon_d$ being replaced with $\varepsilon_d+\sigma B$, while the third one being replaced with $\varepsilon_d+\bar\sigma B$. Then the compensation field $B_c$ can be determined under the condition $\Delta=|\widetilde{\varepsilon}_{d\uparrow}-\widetilde{\varepsilon}_{d\downarrow}|=0$ which means the external field fully compensates the local spin splitting induced by the ferromagnetic Dirac electrons. Figure 6(a) shows that the compensation field $B_c(h)$ as a function of $h$ calculated by the scaling theory is in good agreement with the NRG results calculated via the criterion $n_\uparrow(B_c)=n_\downarrow(B_c)$.

Fixing the magnetic field at $B_c(h)$ for any nonzero $h$, a full Kondo resonance always develops in the local spectral density and its width at half maximum gives the Kondo temperature $T_K$. We find that upon increasing $h$, the Kondo resonance broadens and lowers [Fig.\,6(b)], resulting in an increasing Kondo temperature [Fig.\,6(c)]. This feature is strikingly different from the Kondo physics in conventional ferromagnetic materials. In conventional materials, intensive studies \cite{Martinek2003a, Martinek2003b, Simon2007, Sindel2007, Matsubayashi2007} have demonstrated the destructive influence of ferromagnetism, which reduces $T_K$ and suppresses Kondo correlations, even though the compensation field is already applied. On the contrary, we demonstrate here that although the Kondo effect is absent for a particle-hole symmetric adatom in undoped graphene, the presence of ferromagnetism can induce this effect by enhancing the Kondo temperature and Kondo correlations. This constructive influence of ferromagnetism on the graphene Kondo physics is one of our main findings. An intuitive understanding of this behavior can be achieved by examining the variation of graphene density of states with the ferromagnetic exchange field $h$. As shown in Fig.\,1(a) and Eq.\,(11), the hybridization at the Fermi energy is a monotonically increasing function of $h$, i.e., $\Gamma_\sigma(E_F)=\Gamma_0|h|/D$ for $\mu=\varepsilon_0=0$, so is the graphene density of states. Therefore, as $h$ increases, more Dirac electrons can gather near the Fermi energy, favouring the formation of a Kondo singlet at large energy scales.

\begin{figure}[t]
\includegraphics[width=1.0\columnwidth]{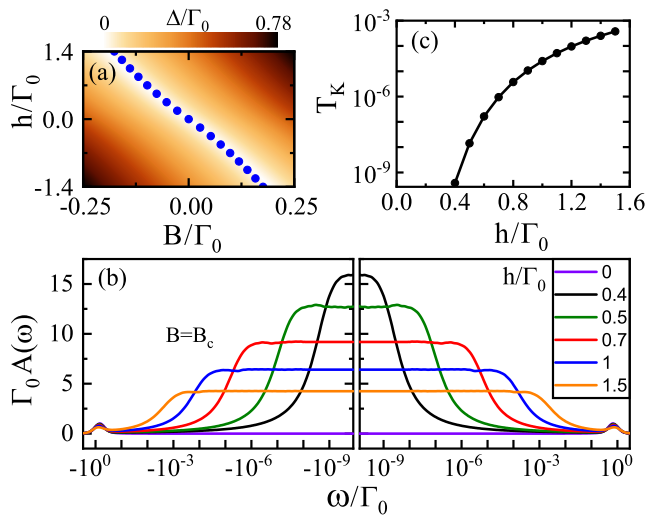}
\caption{(a) Effective spin splitting $\Delta$ of the impurity level as a function of the exchange field $h$ in ferromagnetic graphene and the local external magnetic field $B$, calculated by the perturbative scaling theory. Solid circles are the NRG results of the compensation magnetic field $B_c(h)$. (b) Impurity spectral density $A(\omega)$ at $B=B_c(h)$ for various $h$. (c) Kondo temperature as a function of $h$. Parameters used are $\mu=0$, $\Gamma_0=0.1$, and $\delta=0$.}
\end{figure}

\subsection{C. Kondo physics in the Dirac half-metal regime}
In this subsection, we fix the ferromagnetic exchange field $h/\Gamma_0=0.5$ in graphene and investigate the effect of carrier doping by tuning the chemical potential $\mu$. Note that we only present results of electron doping ($\mu>0$) because for a symmetric ($\delta=0$) impurity the physics arising from hole doping ($\mu<0$) is exactly the same as in the electron doping case. At zero magnetic field, Fig.\,7(a) shows the evolution of Kondo shoulders in the spectral function $A(\omega)=A_\uparrow(\omega)+A_\downarrow(\omega)$ with increasing $\mu>0$. While the shoulder at $\omega>0$, which is of the spin-$\uparrow$ species, monotonically rises as $\mu$ increases, the shoulder at $\omega<0$, which is of the spin-$\downarrow$ species, goes down first and then rises again. Particularly, in the half-metal regime of $\mu=h$ where the Fermi level aligns with the spin-$\uparrow$ Dirac point but $\Gamma_\downarrow(\omega)$ is finite at the Fermi energy [Fig.\,1(b)], the spin-$\downarrow$ shoulder is completely suppressed, leaving only the spin-$\uparrow$ shoulder [see the red curve in Fig.\,7(a)]. Staying in this half-metal regime, we proceed our study by applying the magnetic field $B$ to adjust the local spectral feature. As $B$ increases to compensate the spin asymmetry of local occupations, the spin-$\uparrow$ Kondo shoulder gradually develops into the full Kondo resonance at the Fermi energy [Figs.\,7(b) and 7(c)]. On the other hand, the spin-$\downarrow$ spectral function $A_\downarrow(\omega)$ is always featureless near the Fermi energy, although its two Hubbard bands become more symmetric [Fig.\,7(d)]. Despite the absence of Kondo resonance in $A_\downarrow(\omega)$, the local spin is still fully screened by Dirac electrons in the half-metal regime, since $n_\uparrow=n_\downarrow$ can always be achieved at $B_c$. We find again that the compensation field $B_c(\mu)$ as a function of $\mu$, determined by the NRG method via the criterion $n_\uparrow(B_c)=n_\downarrow(B_c)$, is in good agreement with the corresponding results of the perturbative scaling theory [Fig.\,7(e)].

The Kondo physics revealed here for the Dirac half metals seems peculiar: the Kondo resonance shows up only for the spin-$\uparrow$ component, while $\Gamma_\uparrow(E_F)$ vanishes linearly and $\Gamma_\downarrow(E_F)$ is finite. Nonetheless, this feature is somewhat consistent with previous results obtained in conventional ferromagnetic materials having an energy-independent hybridization $\Gamma_\sigma$. In such impurity systems, the Friedel sum rule \cite{Sindel2007, Matsubayashi2007, Martinek2003b} can be used to relate the spectral function $A_\sigma(E_F)$ at the Fermi energy to the local occupation $n_\sigma$,
\begin{equation}
A_\sigma(E_F)=\frac{\sin^2(\pi n_\sigma)}{\pi\Gamma_\sigma}.
\end{equation}
When equal occupation $n_\uparrow=n_\downarrow$ is achieved by the external compensation field $B_c$, Eq.\,(18) implies that a small hybridization in one spin direction results in a large amplitude of the Kondo resonance in the same spin direction, and vice versa. Although our finding is somewhat understandable from the above analysis, to the best of our knowledge, such kind of fully polarized Kondo resonance appearing only in one spin direction has never been discovered before. Apparently, the exceptional band structure of ferromagnetic graphene is crucial in inducing this unusual phenomenon.

\begin{figure}[t]
\includegraphics[width=1.0\columnwidth]{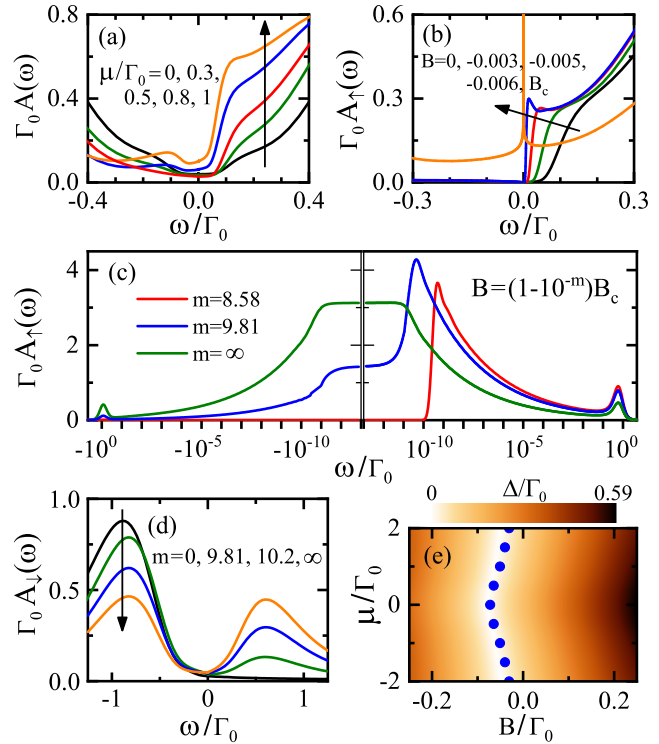}
\caption{(a) Local spectral density $A(\omega)$ for various chemical potentials $\mu$ at zero magnetic field. (b)-(d) Spin-resolved spectral density $A_\sigma(\omega)$ for various magnetic fields $B$ in the half-metal regime ($\mu=h$). Here the compensation field $B_c/\Gamma_0\simeq-0.0646688$ and the exponent $m$ is introduced for the convenience of signifying magnetic fields very close to $B_c$. In (a), (b), and (d), arrows indicate the evolution of spectral functions with the parameters listed in the figures. (e) Effective spin splitting $\Delta$ of the impurity level as a function of $\mu$ and $B$, calculated by the perturbative scaling theory. Solid circles are the NRG results of the compensation magnetic field $B_c(\mu)$. Parameters used are $h/\Gamma_0=0.5$, $\Gamma_0=0.1$, and $\delta=0$.}
\end{figure}

\section{IV. Conclusion}
We have studied the constructive influence of ferromagnetism on the Kondo physics of an impurity atom adsorbed on ferromagnetic graphene. It is demonstrated that for a symmetric impurity in undoped graphene, Kondo correlations can emerge only in the presence of ferromagnetism in graphene and the spin polarization of Dirac electrons strikingly enhances the Kondo temperature. Driving ferromagnetic graphene into the half-metallic regime by carrier doping, we predict an abnormal Kondo resonance that develops in one spin direction but is absent in the opposite direction. These intriguing features can be locally probed by scanning tunneling microscopy, and are in principle also accessible in bulk transport measurements. Our results predicted in this paper, e.g., the fully spin-polarized Kondo resonance, may have potential applications to spintronics based on ferromagnetic graphene.

\section{Acknowledgements}
This work is financially supported by NSF-China (Grants No.\,\,11574007, No.\,\,11504066, No.\,\,11874428, and No.\,\,11874187), National Key Research and Development Program of China (Grant No.\,\,2017YFA0303301), Beijing Municipal Science \& Technology Commission (Grant No.\,\,Z181100004218001), and Innovation-Driven Project of Central South University (Grant No.\,\,2018CX044).

\end{document}